\documentclass[epj]{svjour}
\usepackage{graphics}

\newcommand{\dasgraph}[1]{\centerline{\resizebox{0.75\columnwidth}{!}{\includegraphics{#1}}}}
\newcommand{\DeltaEW}{\ensuremath{\Delta_{\mathrm{EW}}}}
\newcommand{\lambdadB}{\ensuremath{\lambda_{\mathrm{dB}}}}
\newcommand{\kLaser}{\ensuremath{k_{\mathrm{Laser}}}}
\newcommand{\Rb}{$^{85}$Rb}
\newcommand{\ket}[1]{\ensuremath{\left|#1\right\rangle}}

\newcommand{\FmF}[2]{\ket{F=#1,m_F=#2}}
\newcommand{\mean}[1]{\ensuremath{\left\langle#1\right\rangle}}

\begin{document}

\title{An atom interferometer for measuring loss of coherence from an atom mirror}
\author{J. Est\`eve, D. Stevens, C. Aussibal, N. Westbrook, A. Aspect
and C.I. Westbrook}

\institute{Laboratoire Charles Fabry de l'Institut
d'Optique\thanks{\it The Laboratoire Charles Fabry is part of the
Federation LUMAT, FR2764 du CNRS.}, UMR8501 du CNRS, Orsay,
France}

\date{Received: date / Revised version: date}

\abstract{We describe an atom interferometer to study the
coherence of atoms reflected from an evanescent wave mirror. The
interferometer is sensitive to the loss of phase coherence induced
by the defects in the mirror. The results are consistent with and
complementary to recent measurements of specular reflection.
 }

\authorrunning{J. Est\`eve {\it et al.}}

\maketitle

In the past 10 years, atom interferometry has found a number of
applications.
Notable examples are atom gyrometers, gravimeters and accelerometers,
measurements of forward scattering amplitudes for elastic collisions,
and investigations of the Aharanov-Casher effect \cite{berman:1997}.
Here we demonstrate a new application: interferometric
characterization of an atomic mirror.

Using either dipole forces or magnetic fields, it is not difficult
to make a ``mirror'', {\it i.e.} a steep reflecting barrier,
strong enough to reflect atoms with velocities of order 1 m/s, the
velocity acquired in $\sim$ 5 cm of free fall. Both dipole and
magnetic force mirrors can be coupled with a high quality
substrate to guarantee a well defined overall flatness or
curvature, thus giving rise to the evanescent wave mirror
\cite{dowling:1996,balykin:1998}, or to the magnetic mirror
\cite{hinds:1999}. It is now well known however, that a
fundamental difficulty of atomic mirrors is loss of coherence due
to various sources of roughness in the reflecting potential
\cite{landragin:1996a,saba:1999,cognet:1999,lau:1999}. The
extremely small de Broglie wavelength associated with the typical
velocities ($\lambdadB\sim 5$~nm in the case of Rb at 1~m/s),
imposes severe constraints on the small scale roughness of the
substrate -- it must be much better than $\lambdadB /2\pi$
\cite{henkel:1997} before the reflection can be considered
specular, and therefore coherent. This experiment is the first in
which an atomic mirror is used within an interferometer and as
such is the first true demonstration of its coherence.

In a previous paper \cite{savalli:2002}, we reported measurements
of the velocity distribution of atoms from an atom mirror and
measured the fraction of specularly reflected atoms, as well as
the transverse velocity profile of the diffusely reflected ones.
The resolution of this measurement however, was insufficient to
study the lineshape of the specularly reflected distribution -- a
crucial aspect characterizing the effect of the mirror on the
coherence. Here we discuss a related measurement which is able to
focus in more detail on the shape of the specularly reflected
fraction. We have developed an atom interferometer which gives
information complementary to velocity distribution measurements.
We observe fringes whose contrast as a function of path difference
corresponds to the coherence function of the atomic mirror, in
other words to the Fourier transform of the transverse velocity
distribution induced by the mirror. This measurement is
particularly sensitive to the long distance behavior of the
coherence function or to the velocity distribution in the specular
peak, where direct velocity distribution measurements are
impractical. Narrower velocity selection implies fewer atoms and
worse signal to noise. The signal to noise in the interferometric
technique is practically independent of the velocity resolution.
It is the analog of Fourier transform spectroscopy with de Broglie
waves.


\begin{figure}
    \dasgraph{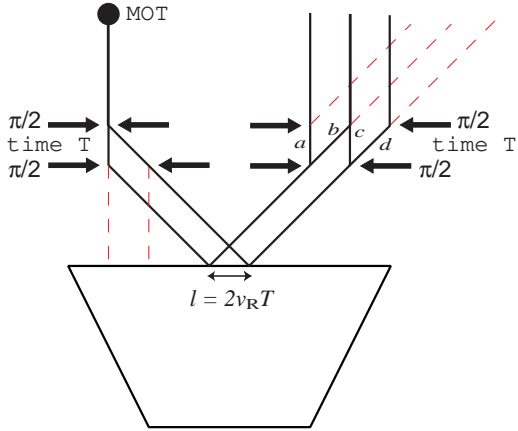}
 \label{Fig:interfere}
 \caption{Diagram of the interferometer. The arrows represent Raman $\pi
 /2$ pulses which create superpositions of different internal states
 and momenta. The atomic mirror is an evanescent wave at the
 surface of a glass prism represented by the trapezoid.
 The dashed lines correspond to paths which are
 eliminated, either during the bounce or during the detection. The
 letters {\it a,b,c} and {\it d}, label the 4 possible paths discussed
 in the text. The path lengths are not a realistic
 representation of the trajectory lengths. }
\end{figure}
A diagram of the experiment is shown in Fig.~1.
Atoms from a MOT are subjected to two $\pi /2$ pulses
which transfer 2 recoil momenta to the atoms.
Their time separation is $T$.
Only one of the internal atomic states is reflected by the mirror as
shown by the solid line paths.
After reflection the two paths are recombined by repeating the Raman
pulse sequence with the same separation time.
The result is an interferometer in which the two possible paths bounce off
different parts of the mirror, separated by $l=2 v_{\rm R} T$, where
$v_{\rm R}$ is the recoil velocity.
By detecting atoms in only one of the two internal states,
interference fringes as a function of the time $T$ are visible as
shown in Fig. 2.
The actual atomic trajectories are parabolic, but we have suppressed
this feature in the figure because the only role played by gravity is
to determine the de Broglie wavelength of the atoms at the moment they
hit the mirror.

The interferometer most resembles one first discussed in Ref.~\cite{borde:1992}
and demonstrated in Ref.~\cite{morinaga:1995}.
The differences here are that
we use 2 photon Raman transitions
rather than 1 photon transitions \cite{kasevich:1991b}
and, more importantly,
that we have placed an
atomic mirror within the interferometer and use the fringes to study
the influence of the mirror on the spatial coherence of the reflected
atoms.
We isolate the effect of the mirror by comparing these fringes to
those obtained by applying all four pulses before the atoms hit the
mirror.
These fringes are also shown in Fig. 2.

It is evident that the mirror strongly reduces the fringe contrast,
and we will discuss the information this reduction gives us below.
But first we will give some experimental details of our setup.
The apparatus is the same as that used in Ref. \cite{savalli:2002},
and the laser pulse sequence is very similar.
We refer the reader to Fig. 1 of that paper for the energy level scheme.
 A \Rb\ MOT is loaded with approximately $10^8$ atoms
in 2~s. The atoms are prepared in the $F=2$ level by turning off
the $2\rightarrow 3$ repumping laser before the trapping beams.
They fall under gravity towards a glass prism 20~mm below.
Starting 8 ms after the atoms begin to fall, two
counter--propagating Raman beams are pulsed on twice for
25~$\mu$s, with a period $T$ between the start of the two pulses.
The two--photon detuning of the first pulse pair is $\delta_1 =
\omega_{a}-\omega_{b}-\omega_{\mathrm{HFS}}$ where $\omega_b$ and
$\omega_a$ are the frequencies of the two Raman lasers,
$\omega_{\mathrm{HFS}}$ is the hyperfine splitting in \Rb,
corrected for the atomic recoils involved in the transition. The
laser parameters are such that the $\pi/2$ condition is fulfilled
for $\delta_1=0$.

A homogeneous magnetic bias field (750~mG), in the propagation
direction of the Raman beams, ensures that only atoms in the
hyperfine sublevel \FmF{2}{0} undergo a transition (into
\FmF{3}{0}). The evanescent wave laser is switched on for 20~ms,
timed to coincide with the arrival of the atoms at the prism, and
a second pair of Raman pulses, separated by the same period $T$,
with two--photon detuning $\delta_2$, is applied. After the 4
pulse sequence, atoms that are not transferred to $F=2$ are
expelled by a 2~ms pulse of light resonant for $F=3$. Finally a
probe beam with repumper is switched on, and the resulting
fluorescence signal is measured with a photomultiplier tube. The
evanescent wave mirror is described in more detail in Ref.
\cite{cognet:1998}. The evanescent wave is red--detuned from
resonance for $F=2$, but blue--detuned for $F=3$. Thus, $F=3$
atoms are reflected. As in Ref. \cite{savalli:2002}, the
evanescent wave detuning $\DeltaEW/2\pi$ is chosen between 500 and
2000~MHz. For more details on the Raman laser setup see
\cite{savalli:2000,esteve:2004}.

Some atoms that do not undergo a transition stimulated by the first Raman pulse
pair can still undergo a transition by spontaneous emission. These atoms make a
background in our data that we measure with a second sequence this time with
$\delta_1$ detuned away from resonance. This measured background is subtracted
from the original signal.
To make observations without the mirror, we proceed in an analogous manner
except that the atoms are initially prepared in the $F=3$ state and a
pushing laser which eliminates the atoms remaining in $F=3$ is
applied between the 2nd and 3rd pulses.
Also, all four Raman pulses, as well as the detection take place before the
atoms hit the mirror.

\begin{figure}
 \dasgraph{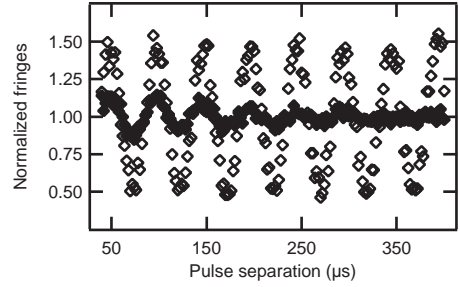}
 \label{Fig:fringe_T}
 \caption{Fringes obtained by scanning the pulse separation $T$
 with (filled circles)
 and without mirror (open circles) for fixed Raman detuning
 $(\delta_{2}-\delta_{1})/2 \pi = 20$~kHz.
 The evanescent wave detuning was 2~GHz.
 }
\end{figure}


To interpret the results, we will analyze the interferometer in
terms of atom interferometry. An equivalent, velocity space
interpretation is possible if one observes that the 1st two Raman
pulses produce two interlaced combs in velocity space
corresponding to the two different atomic internal states
\cite{kasevich:1991a}. The velocity space analysis proceeds by
examining the effect of the mirror on the velocity distribution.
In the interferometric analysis below, we work in position space
and treat the atoms quantum mechanically only along the $x$
(horizontal) direction. The $x$-component of the momentum is
denoted $p$. To treat the effect of the mirror, we use the thin
phase grating approximation \cite{henkel:1994,henkel:1997}, which
supposes that the atomic trajectories are unperturbed by mirror
roughness while within the reflecting potential, and that the
mirror roughness simply adds a phase $\varphi(x) = \frac{4
\pi}{\lambdadB} s(x)$ (this quantity is referred to as
$\delta\varphi(x)$ in Ref.\cite{henkel:1997}) to the matter
wavefront. The quantity $s$ corresponds to the local vertical
deviation of the mirror from a perfectly flat surface.

The atomic source has an rms velocity spread $\delta v$, or
equivalently a coherence length $\xi = \hbar/(m\delta v)$. We will
take the finite coherence length into account by analyzing what
happens to an initial pure state $\ket i$ in the interferometer
and then performing an appropriate, incoherent, average over the
possible states $\ket i$. For example, the initial state could
correspond to a plane wave, in which case the average is over the
velocity distribution of the source, or it could be considered as
a wave packet. The paths in Fig.~1 can be interpreted as the
trajectories of the center of mass of such a wavepacket. Taking
into account the nonzero detuning of the Raman laser results in a
relative phase $\delta_{i} T$ ($i=1$~or~$2$) for each of the
vertical sections of the trajectories between the 1st and 4th
Raman pulses. At the output of the interferometer, after the
fourth $\pi/2$ pulse, the part of the state vector corresponding
to atoms in the $F=2$ state is given by:
\begin{equation}
  \ket{\Psi}=\ket{a}+ \ket{b} + \ket{c} + \ket{d},
\end{equation}
where the letters label the 4 possible paths as in Fig. 1.
If the path separation in the interferometer is greater than the
source coherence length ($l \gg \xi$),
the paths $a$ and $d$
do not contribute to the interference pattern, only to a flat
background.

The amplitudes corresponding to the paths $b$ and $c$ which do interfere can be written:
\begin{eqnarray}
   \ket{b}&=&e^{-i\delta_{1}T}\: U e^{i\varphi(x)} U\ket{i}\\
   \ket{c}&=&e^{-i\delta_{2}T}\: U e^{i\varphi(x+l)} U\ket{i}
\end{eqnarray}
Here $\ket {i}$ is the initial state of the atom and $U=\exp{(-i
\frac{
p^{2}}{2m}\frac{t}{\hbar})}$ is the unitary operator which describes the free
evolution of a state during the time of flight  $t$
from the first $\pi/2$ pulse to the mirror, or from the mirror to the
last $\pi/2$ pulse.

In the above analysis, we have made the approximation that the
duration of the $\pi /2$ pulses can be neglected, compared to all
other time scales in the problem. In our experiment this duration
is not negligible and results in an rms transverse velocity
selection of the atoms by each pulse corresponding to $\delta
v_{\rm Sel}=4.04 /(2 \kLaser \tau)$, where $\tau$ is the duration
of the Blackman pulse and the factor 4.04 converts the total
duration of the Blackman pulse into an rms width ({\it i.e.} a
$1/\sqrt{e}$ half width). For a 25~$\mu $s pulse, this selection
corresponds to a 1~cm/s rms velocity width, or an ``effective''
coherence length of $\xi=80$~nm. In this case the coherence
requirement which permits one to neglect any interference of the
paths $a$ and $d$ is that the path separation $l$ be greater than
this effective coherence length. Taking into account the finite
duration of the pulses also modifies the contrast if the value of
the pulse separation $T$ is close to $\tau$. This effect is very
small but is taken into account in the fits described below. A
more detailed calculation can be found in Ref.\cite{esteve:2004}.

The interference pattern is given by the total probability of finding
the atoms in the $F=2$ state:
\begin{eqnarray}
    \mean{\Psi|\Psi}&=&\mean{a|a} + \mean{b|b} + \mean{c|c} + \mean{d|d}\\
    & &+\mean{b|c} +\mean{c|b}.
\end{eqnarray}
The interference term in the second line is proportional to the
mirror coherence function defined in Ref.\cite{henkel:1997}:
\begin{equation}
 \mathrm{Re}[\mean{b|c}]=
 \mathrm{Re}
 \mean{e^{i\varphi(x+l)-i\varphi(x)}}\cos{(\delta_{2}-\delta_{1})T}
\end{equation}
where \mean{\ldots} refers to a statistical average over the mirror.
The interference pattern is then given by
\begin{equation}
    \mean{\Psi|\Psi}\propto 1+ \frac{1}{2} \mathrm{Re}
    \mean{e^{i\varphi(x+l)-i\varphi(x)}}\cos{(\delta_{2}-\delta_{1})T}
\end{equation}
Fringes are observed either by
varying $T$ or $\delta_{2}-\delta_{1}$ and the coherence function
gives their contrast. A perfect mirror has a coherence
function equal to unity and a fringe contrast of 1/2.

The experiment of Ref.~\cite{savalli:2002}, showed that the
reflected velocity distribution from the mirror is bimodal, having
both a broad (diffuse) and a narrow (specular) component. This
reference identified both mirror roughness and spontaneous
emission as the primary causes of diffuse reflection. We therefore
also expect a bimodal coherence function, consisting of a narrow
part (corresponding to diffuse reflection), and a wide part
corresponding to specular reflection. In terms of the coherence
lengths, Ref.~\cite{savalli:2002} found a length of about 100~nm
for the diffuse part of the coherence function, while only a lower
limit of about 1000~nm (0.1~$v_{\rm R}$) could be given for the
specular part. In the experiment reported here, the minimal time
separation, $T=40$~$\mu$s, corresponds to a path separation of
480~nm in the interferometer. Thus the shape of the diffusely
reflected distribution is inaccessible in the present experiment.
The only effect of diffuse reflection is a loss of contrast, even
for $T=40$~$\mu$s, approximately equal to the fraction $S$ of
specularly reflected atoms. According to Ref.~\cite{savalli:2002},
this fraction is well described by: $S =
\exp\left(-\alpha/\DeltaEW\right)$, where the value of $\alpha$
comes from a fit and is compared to a calculation.

In Fig.~3, we plot the fringe visibility of several runs similar
to those shown in Fig.~2 as a function of the evanescent wave
mirror detuning \DeltaEW. The data were obtained for fixed $T$
(40~$\mu$s) and a varying $\delta_{2}-\delta_{1}$. The fit shown
in the figure corresponds to $\exp\left(-\alpha/\DeltaEW\right)$
times a constant. The data confirm our model of the loss of
contrast and the fit yields $\alpha=$1.5~GHz, in reasonable
agreement with the value found in Ref.~\cite{savalli:2002}. One of
the advantages of the interferometry technique compared to the
simple velocity spectroscopy approach is evident from the error
bars and dispersion of the data in Fig. 3. They are much smaller
than in Fig. 4 of Ref.~\cite{savalli:2002}, and indeed the value
of $\alpha$ deduced here is probably more reliable.

We turn now to the long range behavior of the coherence function.
Fig. 4 shows the behavior of the fringe visibility as a function
of the delay $T$ for a evanescent wave detuning that was fixed at
$\DeltaEW = 2$~GHz. The results show that the contrast decreases
rapidly to zero for delays above 100~$\mu$s. A Gaussian fit to the
data in Fig. 4 gives a $1/\sqrt{e}$ half width of 110~$\mu$s. In
terms of the correlation length, this result corresponds to a
length of 1.3~$\mu$m, just barely longer that the upper limit
established in Ref.~\cite{savalli:2002}. This result was
unexpected.

The loss of contrast can be most easily explained by a small tilt
in the mirror relative to the vertical (Raman velocity
measurements determined it to be 11~mrad). Because of the tilt,
the vertical velocity distribution of the atoms when they hit the
mirror contributes to the horizontal velocity width after the
bounce. The dominant contribution to the vertical velocity
distribution comes from the vertical size of the MOT. An rms MOT
size of 1.7~mm results in a 0.58~mm/s horizontal velocity spread,
and this would account for our observation. The size of the MOT
was not measured during the experiment, but 1.7~mm is plausible.
We have considered other mechanisms for the loss of contrast such
as curvature of the mirror surface, the shape of the waist of the
bouncing laser beam, or diffraction of atoms from the edges. None
of these is large enough to account for the observed loss of
contrast. Note that even if the amount of spontaneous emission
estimated in Ref.~\cite{savalli:2002} is incorrect, it cannot
account for the loss of contrast in Fig. 4 because its intrinsic
length scale $k_{\rm L}^{-1} \sim 100$~nm, ({\it i.e.} of order 1
recoil momentum is imparted to the atoms). The length scale we
observe in Fig. 4 for the loss of contrast is much larger.

\begin{figure} \dasgraph{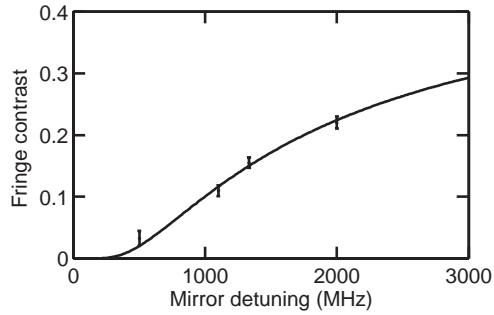} \label{Fig:vis_delta} \caption{Fringe
visibility for $T=40$~$\mu$s as a function of the evanescent wave
detuning \DeltaEW$/2\pi$. The solid curve shows the fitted
function described in the text. } \end{figure}
\begin{figure} \dasgraph{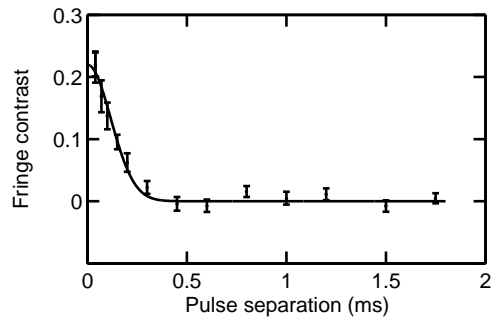} \label{Fig:vis_t} \caption{Visibility of
fringes as a function of pulse separation $T$ for an evanescent
wave detuning of 2~GHz.  The fit is to a Gaussian with a rms width
of 110~$\mu$s.}
\end{figure}

The main conclusion of our work is that, as in traditional optics,
interferometry constitutes an extremely sensitive test of mirror
surface quality. The interferometer is very well compensated for
many parasitic effects effects, such as the incoherence of the source, frequency
fluctuations of the lasers {\it etc.}, as shown by the essentially perfect
contrast observed without the mirror.
Defects in the mirror are readily apparent, and can be easily quantified
as shown by the small error bars in Figs. 3 and 4.
The interferometer is sensitive to coherence lengths much larger than
are accessible to straightforward velocity distribution measurements.

We thank M. Weitz for suggesting this experiment.
This work was supported by the DGA under grant 03.34.003 and by the
European Union under grant IST-2001-38863.

\end{document}